\documentclass[aps,prl,twocolumn,floatfix,showpacs]{revtex4}
\usepackage{graphics}
\usepackage{graphicx}
\usepackage{amssymb}
\usepackage{amsfonts}
\usepackage{amsmath}
\usepackage{bm}
\usepackage{epsf}
\usepackage{version}

\newcommand \be  {\begin{equation}}
\newcommand \ee  {\end{equation}}
\newcommand \bea {\begin{eqnarray} }
\newcommand \eea {\end{eqnarray}}

\newcommand \bd  {\begin{details}}
\newcommand \ed  {\end{details}}

\begin{document}

\title{Kramers degeneracy in a magnetic field and Zeeman 
spin-orbit coupling \\ in antiferromagnetic conductors}
\author{Revaz Ramazashvili}
\affiliation{
ENS, LPTMS, UMR8626, B\^at. 100, 
Universit\'e Paris-Sud, 91405 Orsay, France
}

\date{\today}

\excludeversion{details}


\begin{abstract}
In this article, I study magnetic response of electron 
wavefunctions in a commensurate collinear antiferromagnet. 
I show that, at a special set of momenta, 
hidden anti-unitary symmetry protects Kramers degeneracy 
of Bloch eigenstates against a magnetic field, pointing 
transversely to staggered magnetization. 
Hence a substantial momentum dependence of the transverse 
$g$-factor in the Zeeman term, turning the latter into a 
spin-orbit coupling, that may be present in materials 
from chromium to borocarbides, cuprates, pnictides, 
as well as organic and heavy fermion conductors.
\end{abstract}

\pacs{75.50.Ee}

\maketitle

Antiferromagnetism is a frequent occurrence in materials 
with interesting electron properties: it is found in 
elemental \cite{fawcett_1} and binary \cite{kulikov} 
solids, in numerous borocarbides \cite{mueller}, in 
doped insulators such as cuprates \cite{tranquada}, 
in iron pnictides \cite{cruz}, in various organic 
\cite{chaikin} and heavy fermion \cite{flouquet} 
compounds. The physics of an antiferromagnetic state 
in these materials has been a subject of much research.

\begin{figure}[h]
 \hspace{3cm}
 \epsfxsize=8cm
 \epsfbox{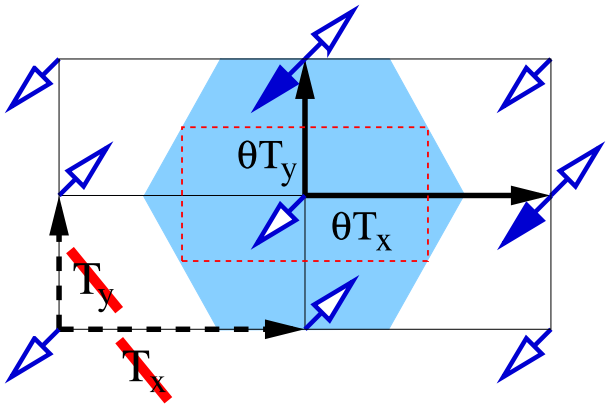}
 \vspace{15pt}
\caption{(color online). 
Doubly commensurate collinear antiferromagnet on a simple 
rectangular lattice. In the absence of magnetism, 
time reversal $\theta$ and primitive translations ${\bf T}_x$ 
and ${\bf T}_y$, shown by dashed arrows, are symmetry operations. 
In the antiferromagnetic state, neither of the three remains 
a symmetry, but the products $\theta {\bf T}_x$ and 
$\theta {\bf T}_y$, shown by solid arrows, do, as illustrated 
by filled spin arrows. Small dashed rectangle at the center 
is the Wigner-Seitz cell boundary in the paramagnetic state, 
while the shaded hexagon is its antiferromagnetic counterpart. 
Notice that neither of the point group operations interchanges 
the two sublattices, hence any point symmetry of the lattice,  
including inversion $\mathcal{I}$, remains a symmetry 
of the antiferromagnetic state. 
} 
\label{fig:AF_real_space}
\end{figure}

In this article, I study response of electron states 
in an antiferromagnet to a weak magnetic field. 
I concentrate on the simplest case: a commensurate 
collinear antiferromagnet, shown schematically in 
Fig. \ref{fig:AF_real_space}, where magnetisation 
density at any point is parallel or antiparallel to 
a single fixed direction ${\bf n}$ of staggered 
magnetisation, and changes sign upon primitive 
translation of the underlying lattice.

In a paramagnet, double degeneracy of single-electron 
eigenstates is usually attributed to symmetry under 
time reversal $\theta$ -- and, indeed, perturbations 
that break $\theta$ (such as ferromagnetism or magnetic 
field) tend to remove the degeneracy. Yet mere violation 
of $\theta$ does not preclude degeneracy: in a commensurate 
centrosymmetric N\'eel antiferromagnet, as in a paramagnet, 
all Bloch eigenstates enjoy Kramers degeneracy 
\cite{herring1} in spite of time reversal being 
broken in the former, but not in the latter.

In an antiferromagnet, staggered magnetisation sets 
a special direction ${\bf n}$ in electron spin space, 
making it anisotropic. Magnetic field along ${\bf n}$ 
removes the degeneracy, as it does in a paramagnet. 
By contrast, in a transverse field, the symmetry remains 
high enough to protect Kramers degeneracy at a special 
set of momenta. Generally, in $d$ spatial dimensions, 
full degeneracy manifold is $(d-1)$-dimensional; 
at its subset, degeneracy is dictated by symmetry. 
This is in marked contrast with a paramagnet, where 
arbitrary magnetic field lifts the degeneracy of all 
Bloch eigenstates.

I show that, at a subset of the degeneracy manifold above, 
it is a hidden anti-unitary symmetry that protects Kramers 
degeneracy of Bloch states in a transverse field. 
The degeneracy gives rise to a peculiar spin-orbit coupling, 
whose emergence and basic properties, along with the 
degeneracy itself, are the main result of this work.

Kramers degeneracy of special Bloch states in a transverse 
field means, that the transverse component $g_\perp$ of 
the electron $g$-tensor vanishes for such states. Not being  
identically equal to zero, $g_\perp$ must, therefore, carry 
substantial momentum dependence, and the Zeeman coupling 
$\mathcal{H}_{ZSO}$ must take the form  
\be 
\label{eq:ZSO}
\mathcal{H}_{ZSO}
 = 
 - \mu_B 
\left[ 
g_{||} 
({\bf H}_\| \cdot \boldsymbol{\sigma}) 
 +
g_\perp ({\bf p}) 
({\bf H}_\perp \cdot \boldsymbol{\sigma})
\right], 
\ee
where 
$
{\bf H}_{||} 
 =
({\bf H} \cdot {\bf n}) 
{\bf n}
$  
and
$
{\bf H}_\perp
 =
 {\bf H}
 - 
{\bf H}_{||} 
$ 
are the longitudinal and the transverse components 
of the magnetic field with respect to unit vector 
${\bf n}$ of staggered magnetisation, 
$
\mu_B 
$ 
is the Bohr magneton, while $g_{||}$ and 
$g_\perp ({\bf p})$ are the longitudinal and 
the transverse components of the $g$-tensor.  

This very momentum dependence of $g_\perp ({\bf p})$  
turns the common Zeeman coupling into a Zeeman spin-orbit 
interaction $\mathcal{H}_{ZSO}$  (\ref{eq:ZSO}), 
whose appearance and key properties are at the focus 
of this work. Zeeman spin-orbit coupling may manifest 
itself spectacularly in a number of ways, which will 
be mentioned below and discussed in detail elsewhere.

Symmetry properties of wave functions in magnetic crystals  
have been studied by Dimmock and Wheeler \cite{dimmock2}, 
who pointed out, among other things, that magnetism not 
only lifts degeneracies by obviously lowering the symmetry, 
but also may introduce new ones. This may happen at the 
magnetic Brillouin zone (MBZ) boundary, under the necessary 
condition that the magnetic unit cell be larger, than the  
paramagnetic one \cite{dimmock2}. 

For a N\'eel antiferromagnet on a square lattice, response 
of electron states to magnetic field was studied in \cite{braluk} 
by symmetry analysis, and in \cite{bralura} within a weak 
coupling model. The present work revisits \cite{braluk}, 
extends it to an arbitrary crystal symmetry and shows, that 
the picture is more rich than envisaged by the authors. 
At the same time, the present work extends \cite{dimmock2} 
by allowing for external magnetic field -- to show how, 
at special momenta, Kramers degeneracy may persist even 
in a transverse magnetic field.

Antiferromagnetic order couples to the electron 
spin $\boldsymbol{\sigma}$ via exchange term 
$({\bf \Delta_r} \cdot \boldsymbol{\sigma})$, 
where ${\bf \Delta_r}$ is proportional to the 
average magnetisation density at point ${\bf r}$. 
Nonzero ${\bf \Delta_r}$ changes sign under time reversal 
$\theta$, and removes the symmetry under primitive 
translations ${\bf T_a}$, thus reducing the symmetry 
with respect to that of paramagnetic state. 
In a doubly commensurate collinear antiferromagnet, 
${\bf \Delta_r}$ changes sign upon ${\bf T_a}$: 
${\bf \Delta_{r+a} = - \Delta_r}$, while 
${\bf T}_{\bf a}^2$ leaves ${\bf \Delta_r}$ intact: 
${\bf \Delta}_{{\bf r}+2{\bf a}} = {\bf \Delta_r}$. 
Even though neither $\theta$ nor ${\bf T_a}$ 
remain a symmetry, their product $\theta {\bf T_a}$ 
does (see Fig. \ref{fig:AF_real_space}).
In a system with inversion center, 
so does $\theta {\bf T_a} \mathcal{I}$, 
where $\mathcal{I}$ is inversion. 

Combined  anti-unitary symmetry 
$\theta {\bf T_a} \mathcal{I}$ 
induces Kramers degeneracy \cite{herring1}: 
If $| {\bf p} \rangle$ is a Bloch eigenstate 
at momentum ${\bf p}$, then 
$\theta {\bf T_a} \mathcal{I} | {\bf p} \rangle$ 
is degenerate with $| {\bf p} \rangle$. 
Since $\theta$ and $\mathcal{I}$ both invert 
the momentum, both $| {\bf p} \rangle$ and 
$\theta {\bf T_a} \mathcal{I} | {\bf p} \rangle$ carry the 
same momentum label ${\bf p}$. 
At the same time, $| {\bf p} \rangle$ and 
$\theta {\bf T_a} \mathcal{I} | {\bf p} \rangle$ 
are orthogonal: recalling that 
$({\bf T_a} \mathcal{I})^2 = - \theta^2 =1$, 
one finds \cite{herring1} 
\be
\label{eq:Kramers}
\langle \bf{p} |
 \theta {\bf T_a} \mathcal{I}
               | \bf{p} \rangle 
 =
 -
\langle \bf{p} |
 \theta {\bf T_a} \mathcal{I} 
               | \bf{p} \rangle. 
\ee
Thus, in spite of broken time reversal symmetry, 
in a centrosymmetric commensurate N\'eel 
antiferromagnet all Bloch states retain 
Kramers degeneracy.

Generally, magnetic field ${\bf H}$ lifts this degeneracy.
However, in a purely transverse field, hidden anti-unitary 
symmetry may protect the degeneracy at a special set 
of points in the Brillouin zone, as I show below. 

In a commensurate collinear antiferromagnet in magnetic 
field ${\bf H}$, electron Hamiltonian has the form  
\be
\label{eq:Hamiltonian_generic}
\mathcal{H} 
 = 
\mathcal{H}_0 + 
({\bf \Delta_r} \cdot \boldsymbol{\sigma}) 
 - ({\bf H} \cdot \boldsymbol{\sigma}), 
\ee
where `paramagnetic' part $\mathcal{H}_0$ is 
invariant under independent action of ${\bf T_a}$ 
and $\theta$, and $g \mu_B$ is set to unity. 
In the absence of the field, all Bloch eigenstates 
of Hamiltonian (\ref{eq:Hamiltonian_generic}) enjoy 
Kramers degeneracy by virtue of Eqn. (\ref{eq:Kramers}).

\begin{figure}[h]
 \hspace{0cm}
 \epsfxsize=5cm
 \epsfbox{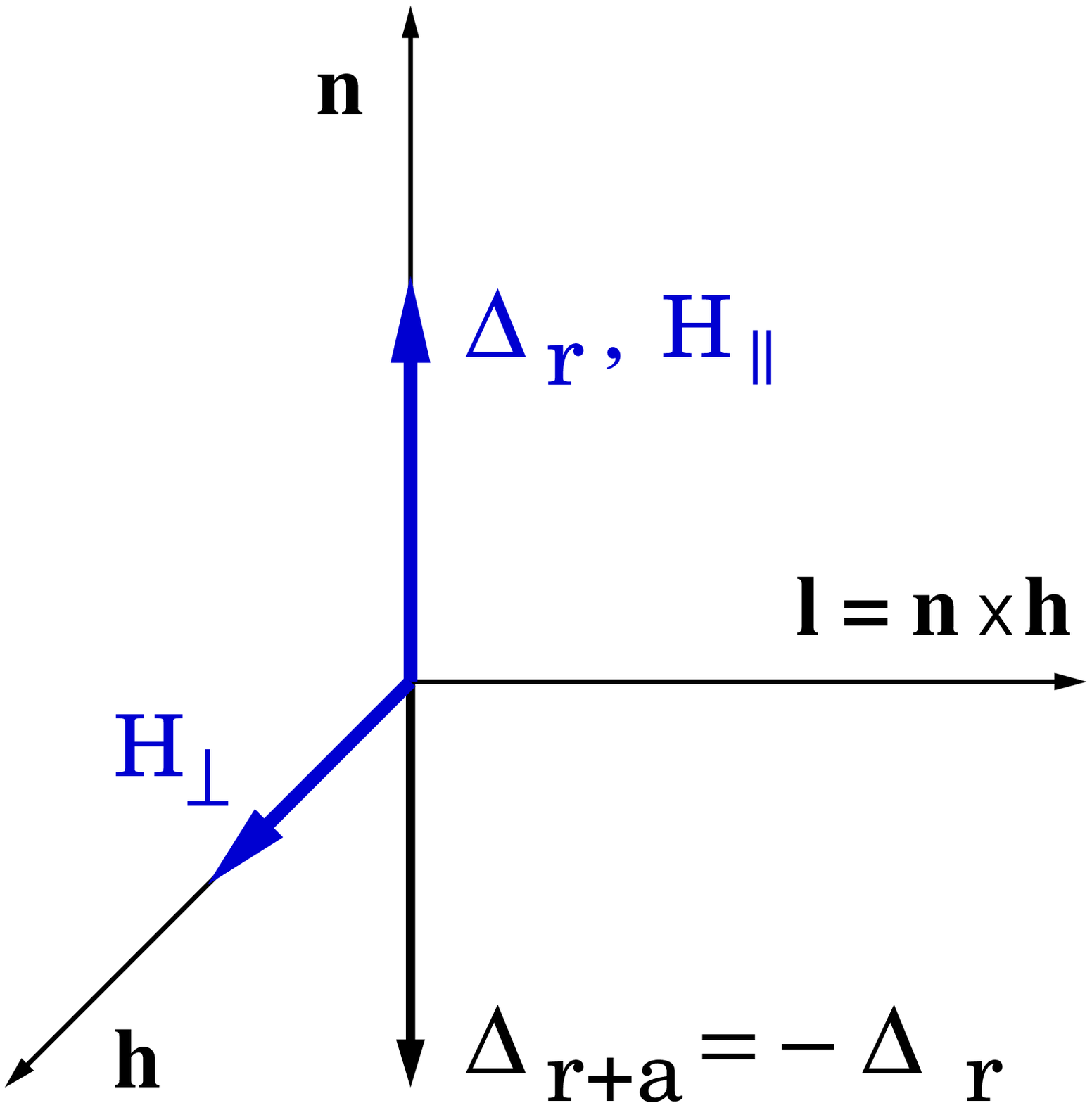}
 \vspace{15pt}
\caption{(color online). 
Relative orientation of ${\bf \Delta_r}$, 
${\bf \Delta_{r+a}}$, ${\bf H}_\|$ and 
${\bf H}_\perp$.
Notice that $\theta$ flips both ${\bf \Delta_r}$ 
and ${\bf H}$, while ${\bf T_a}$ leaves ${\bf H}$ 
intact, but inverts ${\bf \Delta_r}$. 
} 
\label{fig:dhl}
\end{figure}
Consider symmetries of Hamiltonian 
(\ref{eq:Hamiltonian_generic}), involving a combination 
of an elementary translation ${\bf T_a}$, time reversal 
$\theta$, or a spin rotation ${\bf U_m}(\phi)$ around  
axis ${\bf m}$ by angle $\phi$.  
The relative orientation of ${\bf \Delta_r}$, ${\bf H}_\|$ 
and ${\bf H}_\perp$ is shown in Fig. \ref{fig:dhl}. 

Transverse field ${\bf H}_\perp$ breaks the symmetries 
${\bf U_n}(\phi)$ and ${\bf T_a} \theta$ 
(both change ${\bf H}_\perp$), but preserves 
${\bf U_n}(\pi) \theta {\bf T_a}$, their combination 
at $\phi = \pi$. Acting on the exact Bloch state 
$| {\bf p} \rangle$ at momentum ${\bf p}$, 
this combined anti-unitary ope\-ra\-tor 
creates a degenerate partner eigenstate 
${\bf U_n}(\pi) \theta {\bf T_a} | {\bf p} \rangle$, 
which is orthogonal to $| {\bf p} \rangle$ everywhere 
in the Brillouin zone, unless ${\bf p}$ belongs to a 
paramagnetic Brillouin zone boundary:
\be 
\label{eq:thetaTU}
\langle 
{\bf p}
 | 
{\bf U_n}(\pi) \theta {\bf T_a} 
 | {\bf p}  
\rangle 
 = 
e^{-2 i {\bf p \cdot a}}
\langle 
{\bf p}
 | 
{\bf U_n}(\pi) \theta {\bf T_a}
 | {\bf p}  
\rangle. 
\ee
Equation (\ref{eq:thetaTU}) can be derived 
analogously to Eqn. (\ref{eq:Kramers}) 
as soon as one observes, that 
$\left[ {\bf U_n}(\pi) \theta {\bf T_a} \right]^2
 = {\bf T_a}^2$. 

Notice, however, that eigenstate 
$ {\bf U_n}(\pi) \theta {\bf T_a} | {\bf p} \rangle$
carries momentum label $-{\bf p}$ rather than ${\bf p}$. 
Generally, these two momenta are different. However, 
an important exception takes place at the magnetic 
Brillouin zone boundary, if there is a unitary symmetry 
$\mathcal{U}$, transforming $-{\bf p}$ into a momentum, 
equivalent to ${\bf p}$ modulo reciprocal lattice vector 
${\bf Q}$ of the antiferromagnetic state \cite{dimmock2}: 
\be 
\label{eq:Dimmock_equivalence}
- \mathcal{U} {\bf p} = {\bf p} + {\bf Q}.
\ee
In this case, eigenstate 
$\mathcal{U} {\bf U_n} (\pi) \theta {\bf T_a} 
 | {\bf p}  \rangle$ carries momentum label 
${\bf p+Q \equiv p}$, is degenerate with $| {\bf p} \rangle$ 
and orthogonal to it, thus explicitly demonstrating Kramers 
degeneracy at momentum ${\bf p}$ in a transverse field. 
In the simplest case, as in Fig. \ref{fig:BZ_1D} below, 
$\mathcal{U}$ is the unity o\-pe\-ra\-tor. 

Additional insight into the locus of states, whose 
degeneracy persists in a transverse magnetic field, is 
afforded by weak-coupling Hamiltonian of a single electron 
in a doubly commensurate collinear antiferromagnet. 
Let ${\bf Q}$ be the antiferromagnetic ordering 
wave vector (see the examples below); 
${\bf \Delta_r}$ creates a matrix element 
$({\bf \Delta} \cdot \boldsymbol{\sigma})$ between the Bloch 
states at momenta ${\bf p}$ and ${\bf p + Q}$; for simplicity, 
I neglect its possible dependence on ${\bf p}$. 
In magnetic field ${\bf H}$, and at weak coupling, 
Hamiltonian (\ref{eq:Hamiltonian_generic}) 
takes the form \cite{bralura}
\be
\label{eq:weak_coupling_Hamiltonian}
\mathcal{H} 
 = 
\left[
\begin{array}{cc}
\epsilon_{\bf p}
 - ({\bf H} \cdot \boldsymbol{\sigma}) 
 & 
({\bf \Delta} \cdot \boldsymbol{\sigma}) \\
 & \\
({\bf \Delta} \cdot \boldsymbol{\sigma}) 
 & 
\epsilon_{\bf p + Q}
 - ({\bf H} \cdot \boldsymbol{\sigma}) 
\end{array}
\right]
,
\ee
where 
$\epsilon_{\bf p}$  and $\epsilon_{\bf p + Q}$ 
are single-particle energies of $\mathcal{H}_0$ 
in (\ref{eq:Hamiltonian_generic}) at momenta 
${\bf p}$ and ${\bf p + Q}$. 

In a purely transverse field ${\bf H}_\perp$, 
the spectrum of this Hamiltonian is simply 
\be
\label{eq:weak_coupling_spectrum}
\mathcal{E}_{\bf p}
 = 
\eta_{\bf p} 
            \pm
\sqrt{
\Delta^2
 +
\left[
\zeta_{\bf p} \mp (\bf{H}_\perp \cdot \boldsymbol{\sigma})
\right]^2
}, 
\ee 
where 
$
\eta_{\bf p} \equiv 
\frac{\epsilon_{\bf p} + \epsilon_{\bf p + Q}}{2}
$, 
and 
$
\zeta_{\bf p} \equiv 
\frac{\epsilon_{\bf p} - \epsilon_{\bf p + Q}}{2}
$.
Equation (\ref{eq:weak_coupling_spectrum}) illustrates 
several points. Firstly, at half-filling, a gap 
of size $2\Delta$ opens at the chemical potential. 
Secondly, in the absence of magnetic field, each 
eigenstate is indeed doubly degenerate, in agreement 
with the arguments, encapsulated in 
Eqn. (\ref{eq:Kramers}). 
Finally, Eqn. (\ref{eq:weak_coupling_spectrum}) shows, 
that the degeneracy persists in a transverse field 
(and, therefore, $g_\perp({\bf p})$ 
in Eqn. (\ref{eq:ZSO}) vanishes) 
whenever $\zeta_{\bf p} = 0$. Barring a special situation, 
this equation defines a surface in three dimensions, a line 
in two, and a set of points in one. Furthermore, as shown 
above, this manifold must contain all points, satisfying 
Eqn. (\ref{eq:Dimmock_equivalence}).

Notice that transverse magnetic field not only introduces 
the last term in Hamiltonian (\ref{eq:Hamiltonian_generic}), 
but also tilts the sublattices. However, the resulting 
magnetisation has the same symmetry as the field, 
and thus does not remove the degeneracy. 

Consider examples. 
In one dimension, magnetic Brillouin zone boundary 
reduces to two points ${\bf p} = \pm \frac{\pi}{2a}$, 
which in fact coincide modulo antiferromagnetic wave 
vector ${\bf Q} = \frac{\pi}{a}$, that is also the 
reciprocal lattice vector of the antiferromagnetic 
state (see Fig. \ref{fig:BZ_1D}). In terms of the 
general condition (\ref{eq:Dimmock_equivalence}), 
this is the simplest case: $\mathcal{U} = {\bf 1}$. 

As a result, at ${\bf p} = \pm \frac{\pi}{2a}$, 
the two exact Bloch states in a transverse field, 
$| {\bf p} \rangle$ and 
$\theta {\bf T_a U_n} (\pi) | {\bf p} \rangle$, 
correspond to the {\em same} momentum ${\bf p}$, and 
are degenerate by virtue of $\theta {\bf T_a U_n} (\pi)$ 
being a symmetry. Equation (\ref{eq:thetaTU}) 
guarantees their orthogonality, thus protecting 
Kramers degeneracy at momentum ${\bf p} = \pm \frac{\pi}{2a}$ 
against transverse magnetic field. 
\begin{figure}[h]
 \hspace{-0.5cm}
 \epsfxsize=4cm
 \epsfbox{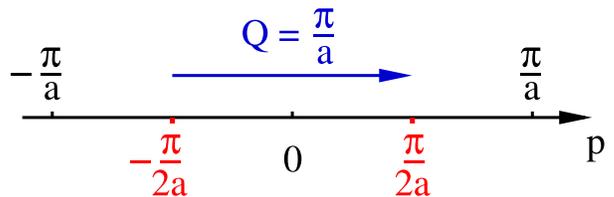}
 \vspace{15pt}
\caption{(color online).  
The paramagnetic (${\bf p} = \pm\frac{\pi}{a}$), and 
the antiferromagnetic (${\bf p} = \pm\frac{\pi}{2a}$) 
Brillouin zone boundaries of a one-dimensional N\'eel 
antiferromagnet. 
In the antiferromagnetic state, the two points 
${\bf p} = \pm\frac{\pi}{2a}$ are identical modulo 
the antiferromagnetic reciprocal lattice vector 
${\bf Q} = \frac{\pi}{a}$. At these two points, 
anti-unitary symmetry ${\bf U_n}(\pi){\bf T_a} \theta$ 
protects Kramers degeneracy against transverse 
magnetic field. 
} 
\label{fig:BZ_1D}
\end{figure}

\begin{figure}[h]
 \hspace{3cm}
 \epsfxsize=8cm
 \epsfbox{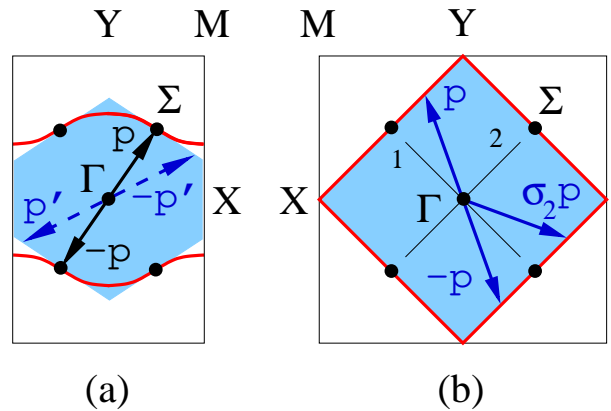}
 \vspace{15pt}
\caption{(color online).  
Geometry of the problem.
(a) The Brillouin zone for a simple rectangular 
lattice (the rectangle), and its antiferromagnetic 
counterpart (MBZ, shaded hexagon). Thick curve (red 
online), passing through point $\Sigma$, shows a 
typical degeneracy line $g_\perp ({\bf p}) = 0$. 
At the MBZ boundary, only momentum ${\bf p}$ 
at point $\Sigma$ is equivalent to ${- \bf p}$ 
modulo the reciprocal lattice vector of the 
antiferromagnetic state. For a generic ${\bf p'}$, 
shown by the dashed arrow, this is no longer true. 
(b) The Brillouin zone of a simple square lattice 
and  its antiferromagnetic counterpart (shaded 
diagonal square). The degeneracy line must contain 
the entire MBZ boundary, shown in red online. 
} 
\label{fig:BZ_rectangular}
\end{figure}

Now, consider a two-dimensional antiferromagnet 
of simple rectangular symmetry, with the ordering 
wave vector ${\bf Q}=(\pi,\pi)$, as shown in 
Fig.\ref{fig:AF_real_space}. 
In a transverse magnetic field, degeneracy persists 
at a line in the Brillouin zone, by virtue of 
Eqn. (\ref{eq:weak_coupling_spectrum}). 
I will show that, in the rectangular case, the 
degeneracy line must contain point $\Sigma$ 
(i.e. the star of point ${\bf p} = {\bf Q}/2$) 
at the center of the magnetic Brillouin zone 
boundary (see Fig. \ref{fig:BZ_rectangular}(a)). 
Consider a Bloch state $| {\bf p} \rangle$ 
at momentum ${\bf p}$ in a transverse field. 
As discussed above, eigenstate 
$ \theta {\bf T_a U_n}(\pi) | {\bf p} \rangle$ at 
momentum $-{\bf p}$ is degenerate with $| {\bf p} \rangle$ 
and, according to Eqn. (\ref{eq:thetaTU}), must be 
orthogonal to it unless $({\bf p} \cdot {\bf a})$ 
is an integer multiple of $\pi$. At points $\Sigma$ 
(the star of ${\bf p} = {\bf Q}/2$), X, and Y, 
momenta ${\bf p}$ and $-{\bf p}$ coincide modulo 
a reciprocal lattice vector of the antiferromagnetic 
state. However, at points X and Y (as well as at the 
entire vertical segment of the MBZ boundary in Fig. \ref{fig:BZ_rectangular}(a)), 
$({\bf p} \cdot {\bf a})$ is an integer multiple 
of $\pi$; hence $| {\bf p} \rangle$ and 
$\theta {\bf T_a U_n}(\pi) | {\bf p} \rangle$ 
are not obliged to be orthogonal  there as per 
Eqn. (\ref{eq:thetaTU}). Thus, $\Sigma$ 
is the only point at the MBZ boundary, where the 
two degenerate states $| {\bf p} \rangle$ and 
$ \theta {\bf T_a U_n}(\pi) | {\bf p} \rangle$ 
are orthogonal and correspond to the same momentum. 
Dashed arrows in figure \ref{fig:BZ_rectangular}(a) show, 
that, for a generic point ${\bf p}'$ at the MBZ boundary, 
no symmetry operation relates $-{\bf p}'$ to a vector, 
equivalent to ${\bf p}'$. Hence it is only at point 
$\Sigma$, that the symmetry protects Kramers 
degeneracy against transverse magnetic field. 
As in the one-dimensional example above, in terms of 
Eqn. (\ref{eq:Dimmock_equivalence}) this corresponds 
to the simplest case of $\mathcal{U}={\bf 1}$.

Promotion from rectangular to square symmetry 
brings along invariance under reflections $\sigma_{1, 2}$ 
in either of the two diagonal axes $1$ and $2$, 
passing through point $\Gamma$ in 
Fig. \ref{fig:BZ_rectangular}(b). 
As a result, eigenstate 
$\sigma_1 \theta {\bf T_a U_n}(\pi) | {\bf p} \rangle$ 
at momentum $\sigma_2 {\bf p}$ 
(Fig. \ref{fig:BZ_rectangular}(b)) is also degenerate 
with $| {\bf p} \rangle$ and orthogonal to it, as one 
can show analogously to the examples above. In terms 
of general condition (\ref{eq:Dimmock_equivalence}), 
this means $\mathcal{U} = \sigma_{1, 2}$.

For momentum ${\bf p}$ at the MBZ boundary 
in Fig. \ref{fig:BZ_rectangular}(b), ${\bf p}$ 
and $\sigma_2 {\bf p}$ differ by a reciprocal 
lattice vector and thus coincide. Hence, for a 
square-symmetry lattice in a transverse field, 
Kramers degeneracy is protected by symmetry at 
the entire MBZ boundary, as shown in Fig. 
\ref{fig:BZ_rectangular}(b). 

Degeneracy of special Bloch states in a transverse field 
hinges only on the symmetry of the antiferromagnetic state, 
and thus holds equally in a strongly correlated or a weakly 
coupled material -- provided long-range antiferromagnetic 
order and well-defined electron quasiparticles. Under these 
conditions, quantum fluctuations of the antiferromagnetic 
order primarily renormalize the sublattice magnetisation, 
but leave intact the degeneracy of special electron states 
in a transverse field -- certainly in the leading order 
in fluctuations.

Now, $g_\perp ({\bf p})$ can be expanded in a vicinity 
of the degeneracy line  $g_\perp ({\bf p}) = 0$. With the 
exception of higher-symmetry points, such as point $X$ 
in Fig. \ref{fig:BZ_rectangular}(b), the leading term of the 
expansion is linear in momentum deviation ${\bf \delta p}$ 
from the degeneracy line: 
\be
g_\perp ({\bf p})
   \approx 
\frac{{\bf \Xi}_{\bf p} \cdot \delta {\bf p}}{\hbar}, 
\label{eq:expand_g}
\ee 
where ${\bf \Xi}_{\bf p}/\hbar$ is momentum gradient of 
$g_\perp ({\bf p})$ at point ${\bf p}$ on the degeneracy 
line. Inversion symmetry requires, that ${\bf \Xi_{\bf p}}$ 
be a pseudo-vector and change sign upon inversion, 
and Eqn. (\ref{eq:weak_coupling_spectrum}) shows, 
that $\Xi_{\bf p}$ is of the order of the antiferromagnetic 
coherence length $\xi \sim \frac{\hbar v_F}{\Delta}$.

Zeeman spin-orbit coupling (\ref{eq:ZSO}) induces 
a number of interesting effects. For instance, 
substantial momentum dependence of $g_\perp ({\bf p})$ 
means, that the Electron Spin Resonance (ESR) frequency 
of a carrier in a vicinity of the degeneracy line varies 
along the quasiclassical trajectory. In a weakly doped 
antiferromagnetic insulator, this means inherent 
broadening of the ESR line with doping and, eventually, 
loss of the ESR signal. In fact, this may well be the 
reason behind the long-known `ESR silence' 
\cite{shengelaya} of the cuprates. Suppression of 
transverse Pauli susceptibility is another simple 
consequence of vanishing $g_\perp({\bf p})$. 

Finally, momentum dependence of $g_\perp({\bf p})$ 
allows excitation of spin flip transitions by AC 
{\em electric} rather than magnetic field \cite{rr}
 -- a vivid effect of great promise for controlled 
spin manipulation, so much sought after in 
spin electronics. Its absorption matrix elements 
are defined by $\Xi_{\bf p} \sim \xi$ of Eqn. 
(\ref{eq:expand_g}), and exceed those of ESR 
at least by two orders of magnitude.  
According to Eqn. (\ref{eq:expand_g}), 
resonance absorption in this phenomenon shows 
a non-trivial dependence on the orientation 
of the AC electric field with respect to the 
crystal axes, and on the orientation of the 
DC magnetic field with respect to staggered 
magnetisation. 

I thank LPTMS for kind hospitality, 
and IFRAF for generous support.

\end{document}